\documentclass[aps,pra,twocolumn,floatfix,superscriptaddress,showpacs]{revtex4}
\usepackage{amssymb}   
\usepackage{graphicx}
\usepackage{color,epsfig}
\usepackage{amsmath}
\usepackage{bm}        
\usepackage{multirow}
\usepackage{booktabs}
\usepackage{pstricks} 
\usepackage[latin1]{inputenc}
\usepackage{color}

\begin{document}

\title{Beyond universality: parametrizing ultracold complex-mediated
reactions using statistical assumptions.}

\author{Manuel Lara\footnote{Corresponding author. E-mail:
 {\em manuel.lara@uam.es}}}
\affiliation{ Departamento de Qu\'imica F\'isica Aplicada, Facultad
de Ciencias, Universidad Aut\'onoma de Madrid, 28049 Madrid, Spain}

\author{P. G. Jambrina}
\affiliation{Departamento de Qu\'imica F\'isica, Facultad
de Qu\'imica, Universidad Complutense, 28040 Madrid,
Spain}

\author{J.-M. Launay} \affiliation{Institut de Physique de Rennes, UMR CNRS
6251, Universit\'e de Rennes I, F-35042 Rennes, France}

\author{F. J. Aoiz\footnote{Corresponding author. E-mail: {\em aoiz@quim.ucm.es}}}
\affiliation{Departamento de Qu\'imica F\'isica, Facultad de Qu\'imica, Universidad Complutense, 28040 Madrid, Spain}

\date{\today}

\begin{abstract}
We have calculated accurate quantum reactive and elastic
cross-sections for the prototypical barrierless reaction
D$^{+}$ + H$_2$($v$=0, $j$=0) using the hyperspherical
scattering method. The considered kinetic energy ranges
from the ultracold to the Langevin regimes. The
availability of accurate results for this system allows to
test the quantum theory by Jachymski {\em et al.}
[\textit{Phys. Rev. Lett.} \textbf{110}, 213202 (2013)] in
a nonuniversal case. The short range reaction probability
is rationalized using statistical model assumptions and
related to a statistical factor. This provides a means to
estimate one of the parameters that characterizes ultracold
processes from first principles. Possible limitations of
the statistical model are considered.
\end{abstract}

\maketitle

The increasing availability of cold and ultracold samples
of atoms and molecules has sprung a great interest in
chemical reactions at very low
temperatures~\cite{Ospel,Staanum,Zahzam,hudsonexp,Nar1}.
Although new experimental approaches \cite{Nar1} appear
highly promising, advances in the field are hampered by
technical problems to produce most molecules at low
temperatures and high enough densities. In contrast to
neutral species, ions can be easily trapped and cooled. The
technology of Coulomb crystals in radio frequency ion traps
~\cite{Coulomb}, and the possibility of combining them with
traps for neutrals or with slow molecular
beams~\cite{hybrid2,hybrid3}, anticipate a great progress
in the analysis of ion-neutral reactions in the nearly
future.

Theoretical simulations employing standard {\em ab initio}
approaches are not feasible for most of the systems thus
far considered. For heavy systems (more convenient
experimentally) there are not potential energy surfaces
(PES) accurate enough to describe processes near
thresholds. Additionally, most of exact dynamical
treatments face insurmountable problems in such regime.
However, in contrast to short-range (SR) chemical
interactions, those occurring at long-range (LR) can be
more easily calculated. Moreover, theoretical approaches
based on the only knowledge of the LR part of the PES have
been able to describe recent experimental findings nearly
quantitatively~\cite{Ospel,Jach:02}. Indeed, processes at
very low collision energies privilege LR interactions,
leading to the idea of {\em universality} in extreme
cases~\cite{Fara}: the result of the collision depends
exclusively on the LR behavior and not on the details of
the PES. In this regard, recently proposed LR
parametrization procedures
\cite{Koto2012,Jach:02,Gao2008,QueB} are very appealing.
Using experimental data, these models are able to predict
non-measured values providing some insight into the
underlying interactions. In particular, the approach in
based on multichannel quantum-defect theory (MQDT),
provides analytical expressions which can be easily
compared with experimental data
~\cite{Jach:02,Jach:03,Jach:04}. The model has been
recently applied to a variety of systems
\cite{Jach:02,Oster14,Nar5}. In particular, for the Penning
ionization of Ar by He($^{3}$S)~\cite{Nar1}, the rate
coefficients have been fitted in a wide range of collision
energies using only two parameters \cite{Jach:02}. However,
the parameters of the model are phenomenological and they
had not been determined before from first principles.

In this work, we present accurate calculations for the
reactive collision D$^{+}$ + H$_2$($v$=0, $j$=0) using the
hyperspherical reactive scattering method
\cite{Launayfirst}. We consider collision energies which
range from the ultracold regime, where only one partial
wave is open, to the Langevin regime where many of them
contribute. These calculations allow to test the quantum
model by Jachymsky {\em et al.}~\cite{Jach:02,Jach:03} by
comparison with accurate theoretical results in a realistic
atom+diatom system, providing a way to estimate one of the
parameters using statistical model assumptions.

\begin{figure}[t]
 \begin{center}
\includegraphics[width=50ex]{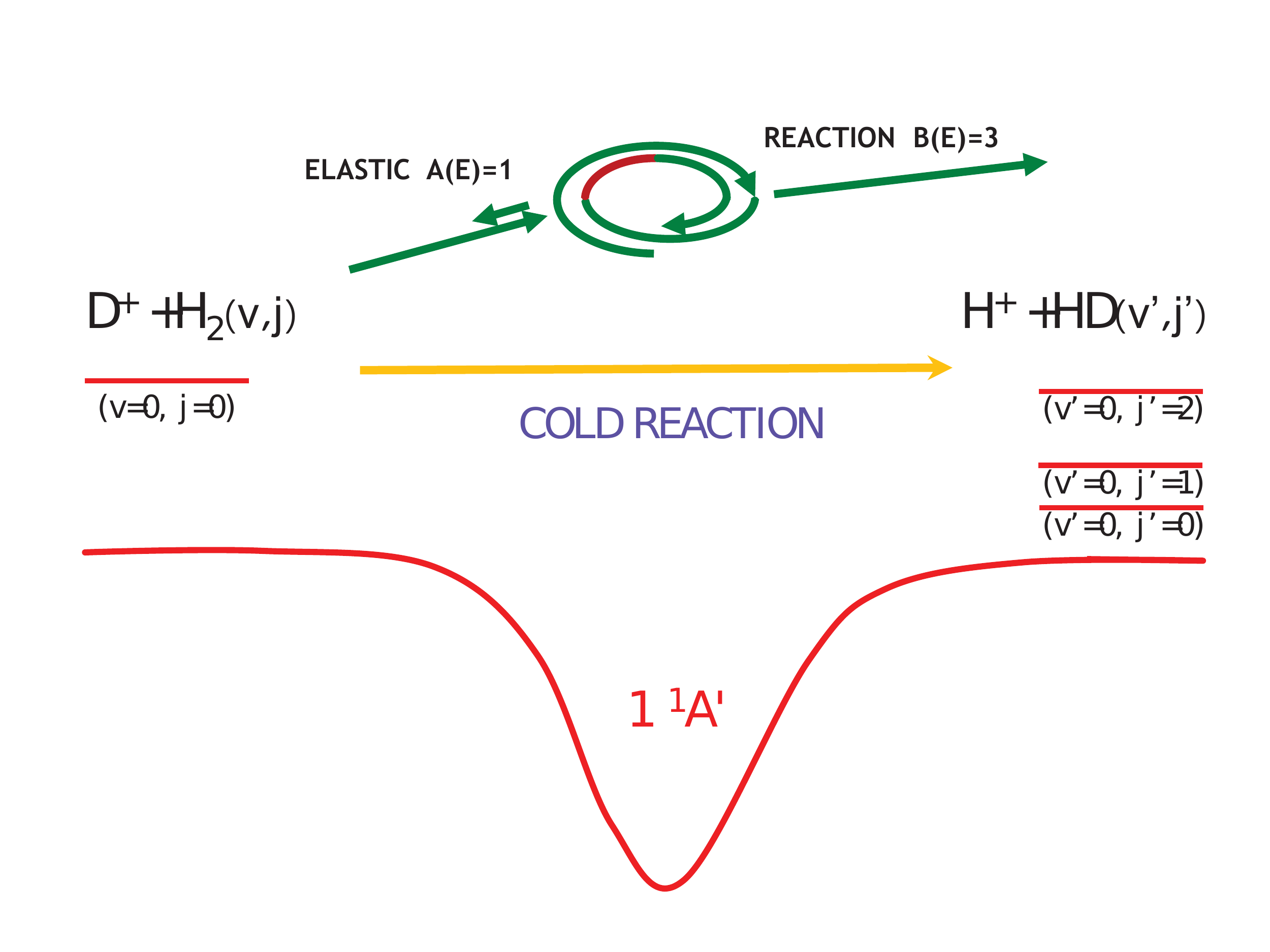}
\caption{Sketch of the intrinsic reaction path and the
rovibrational states involved in the reaction at the studied energies. $A(E)$ and $B(E)$ are the number
of incoming and outgoing channels, respectively (1 and 3 for $J$=0).}
\label{fig1} \vspace{-0.5cm}
\end{center}
\end{figure}

The H$^{+}$+H$_2$ system (and isotopic variants) is the
prototype of ion-molecule reactions, which are usually
nearly barrierless and exhibit large cross sections due to
their LR, $\propto - C_{n}/R^{n}$, $n$=4, potentials. The
process occurs via the strongly bound intermediate
H$_{3}^{+}$, the most abundant triatomic ionic species in
dense interstellar clouds \cite{MGHO:AJ99}. At energies
below $\approx$ 1.7 eV, the proton exchange is the only
reactive channel, and the process can be described on the
ground adiabatic PES
\cite{CGRHLBABTW:JCP08,JAAHS:PCCP10,Honv2,Honverr}. Since
the PES is characterized by a deep well ($\approx 4.5$ eV),
as illustrated in Fig.~\ref{fig1}, rigorous statistical
models \cite{Rack0,AGS:JCP07} have been applied to this
reaction in the low and thermal energy regimes
\cite{AGS:JCP07,JAAHS:PCCP10,GHJAL:JCP09,Honv1,LSH-JPCA2014}
in good agreement with accurate calculations. Specifically,
the D$^+$+H$_2$$\rightarrow$ H$^+$ + HD reaction features a
small exoergicity (difference of zero point energies).
Experiments to determine state specific rate coefficients
at energies as low as 12~K \cite{G:ACP92} have been carried
out, and lower temperatures are expected to be feasible
soon \cite{Gerl}.

The deep ultracold regime, governed by Wigner
laws~\cite{Sade}, is described in terms of the scattering
length. The latter largely varies with slight changes of
the interaction potential whenever a bound state occurs in
the vicinity of the threshold. Only for very particular
atoms~\cite{Przy,Vassen,Knoop}, it has been possible to
determine theoretically the experimental scattering length,
so we cannot expect {\em ab initio} methodology to have
predictive power at ultracold energies in this atom+diatom
system. In the spirit of the work by Gribakin {\em et al.}
\cite{Griba}, we can consider our study as an effort to
determine a ``characteristic'' scattering length. Besides,
assuming that the interaction of the system is reasonably
described by the current PES, we can use our results to
test recent methodologies, like the approach in
Refs.~\cite{Jach:02,Jach:03}.

In the (ultra-)cold regime both accurate descriptions of
the LR interactions and dynamical propagations up to very
large distances are strict requirements. The PES by Velilla
{\em et al.}~\cite{VLABP:JCP08}, which includes the LR
interactions in the functional form, satisfies the first
requirement. The dominant contributions involve the
charge-quadrupole , $\propto - R^{-3}$, and the
charge-induced dipole, $\propto - R^{-4}$, interactions.
However, only the latter contributes to collisions in $j=0$
\footnote{The integral $\langle j=0 | P_2 | j=0 \rangle$ is
null, and the contributions from $\sim R^{-3}$ and
anisotropic polarization terms vanish. \cite{VLABP:JCP08}}.
The second requirement is fully satisfied by the
hyperspherical quantum reactive scattering method developed
by Launay {\em et al.}~\cite{Launayfirst,hon04}, recently
modified in order to allow the accurate inclusion of LR
interactions \cite{Sol02,Lara2}. These modifications are
used here for the first time in order to converge
scattering results at ultracold energies in a system where
the $\propto - R^{-4}$ behavior implies propagations up to
hundreds of thousands of a.u. The logarithmic derivative
(LD) in hypersperical coordinates is propagated in the SR
region up to a particular value of the hyper-radius and
matched to a set of LR functions. As the kinetic energy in
the SR region is very large ($\sim$ eV), the LD is almost
constant for changes in the collision energy on the order
of mK. We can calculate the LD for a few energies, and use
interpolations to obtain it for the rest, thus reducing the
computational cost\cite{Bohn07}

\begin{figure}
 \begin{center}
\includegraphics[width=60ex]{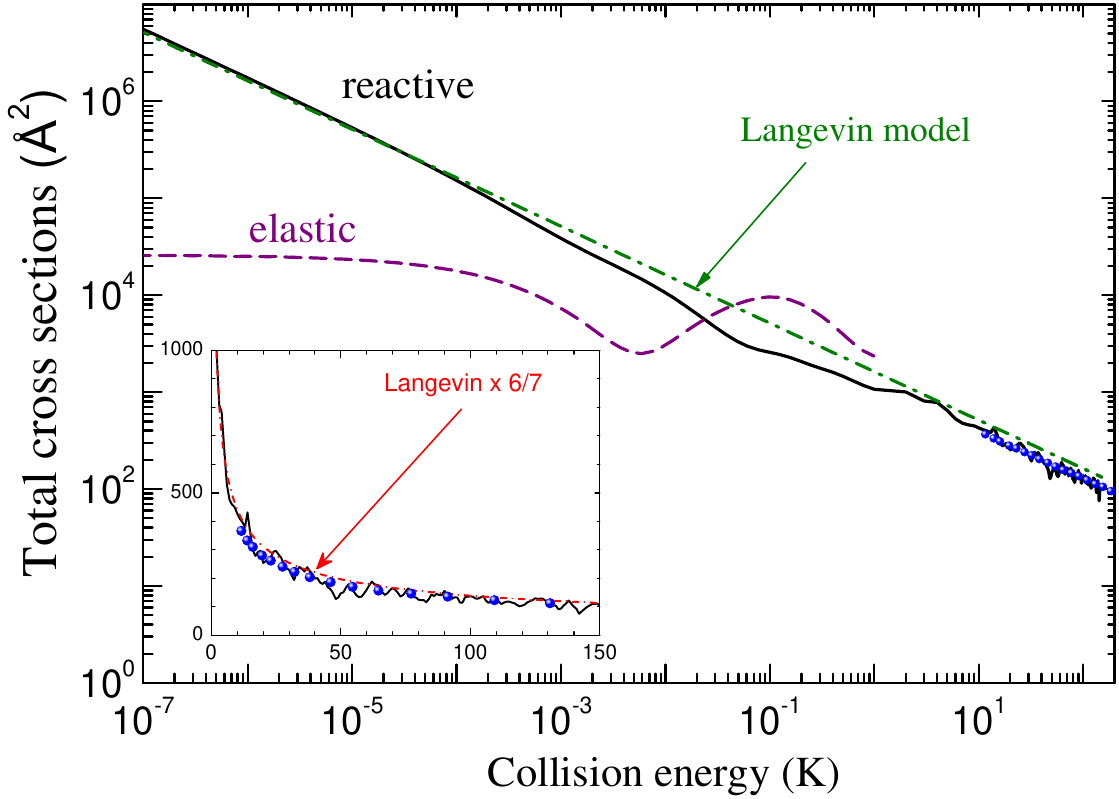}
\caption{Calculated reaction and elastic total cross-section for the collision
D$^{+}$+H$_2$ ($v$=0, $j$=0) compared with the Langevin prediction.
The inset shows the comparison of the reaction cross-sections
with the experimental data \cite{G:ACP92} and
the result of the statistically corrected
Langevin model.} \label{fig2}
\vspace{-0.5cm}
\end{center}
\end{figure}

The reaction cross-section in the 10$^{-7}$--150\,K energy
range is plotted in Fig.~\ref{fig2} and compared with the
Langevin model (LM) expression for the cross-section,
$\sigma_{\rm L}(E)= 2 \pi (C_4 /E)^{1/2}$. The LM is
commonly used to rationalize collisions in the regime where
many partial waves are open. We have determined the value
$C_4$=2.71 a.u. using the effective potential as a function
of $R$ which results from averaging the PES $V(R,r,\theta)$
over $r$ and $\theta$ Jacobi coordinates, with the ($v$=0,
$j$=0) probability distribution. Since the reaction is
mediated by a long-lived complex, the collision process can
be split in the step of the complex formation (capture),
and that of its decomposition. Under this assumption, the
LM implies that, once the centrifugal barrier is overcome,
the system is captured in the complex which subsequently
decomposes into the H$^{+}$+HD arrangement channel with
unit probability. The LM regime can be associated with the
high energy part of the plot, with 5 and 17 partial waves
open at 1\,K and 150\,K, respectively. The calculated
cross-sections are found smaller than the LM prediction in
this energy range. This comes to no surprise; actually,
only a fraction of the complexes decomposes into the
products arrangement and hence the LM hypothesis does not
hold.

We can improve the LM using statistical model arguments. In
complex mediated reactions, the statistical {\em ansatz},
$P^J_{r}(E) \approx  P^J_{\rm capt}(E) \times P_{\to {\rm
prod}}(E)$, can be applied, where $ P^J_{r}(E)$ is the
reaction probability of a given initial rovibrational state
and total angular momentum, $J$ (orbital, $l$, plus
rotational, $j$), $P^J_{\rm capt}(E)$ the probability for
the reagents to be captured in the complex, and $P_{\to
{\rm prod}}(E)$ is the {\em statistical factor}\,: the
probability of emerging into the product arrangement
channel when the complex decomposes. If there is a complete
randomization of the energy within the complex (ergodic
hypothesis), the statistical factor will be independent of
the initial state of the reagents, only subject to
conservation of energy, $J$, and parity. Roughly speaking,
the fraction of complexes which decompose into the
reactants (D$^{+}$+H$_2$) or products (H$^+$+HD) is
proportional to the respective number of scattering
channels energetically available, denoted with $A(E)$ and
$B(E)$ respectively, considering all of them as
equiprobable. Accordingly, the statistical factor can be
approximated by  $P_{\to {\rm prod}}(E) = B(E)/[A(E)+B(E)]$
(number of favorable outcomes over the total number of
equiprobable outcomes).  At the considered energies, only
three HD rovibrational states are open, as shown in
Fig.~\ref{fig1}, and for $J>1$ we find that  $A(E)=1$ and
$B(E)=6$, and $P_{ \to {\rm prod}}=6/7$ ($\approx$ 86\,\%)
~\footnote[4]{As an example, for $J=2$ the incoming channel
has quantum numbers ($v$=0, $j$=0, $l$=2, $J$=2). Only
those product channels, ($v'$, $j'$, $l'$, $J$), with the
same total angular momentum $J=2$, and parity
$\epsilon=(-1)^{j+l}=+1$, are coupled to the incoming one.
Considering all the possible combinations, we get that
$A(E)=1$ and $B(E)=6$.}. For $J=0$ and $J=1$ the
statistical factors are 3/4 (75\%) and 5/6($\approx$ 83\%),
respectively.  If the collision energy is high enough for
many partial waves to contribute, $P_{ \to {\rm prod}}(E)
\approx 6/7$, and $\sigma_{\rm r}(E) \approx P_{ \to {\rm
prod}} \cdot \sigma_{\rm L}(E)$. Therefore, 6/7 appears
naturally as a statistical factor to correct the LM
expression. When the number of product channels is large
enough, $B>>A$, then $P_{ \to {\rm prod}}\approx 1$ and the
result is $\sigma_{\rm L}(E)$. More accurate statistical
implementations, which evaluate $A(E)$ and $B(E)$ as
capture probabilities~\cite{Rack0,AGS:JCP07}, lead to
similar conclusions.

The inset of Fig.~\ref{fig2}, compares the calculated
reaction cross-section with the experimental data from
Ref.~\cite{G:ACP92}. The corrected LM result is also shown
and found in a very good agreement with both the
experiment and the present calculations. The similarity of
the theoretical results with the experiment is remarkable
considering that the latter was performed with $n$-H$_2$.
This is not coincidental: according to  recent
calculations in the high energy range ($>100$\,K)
\cite{Honv2}, cross-section of $j$=0 and $j$=1 are similar.
\begin{figure}
 \begin{center}
\hspace{-0.5cm}
\includegraphics[width=65ex]{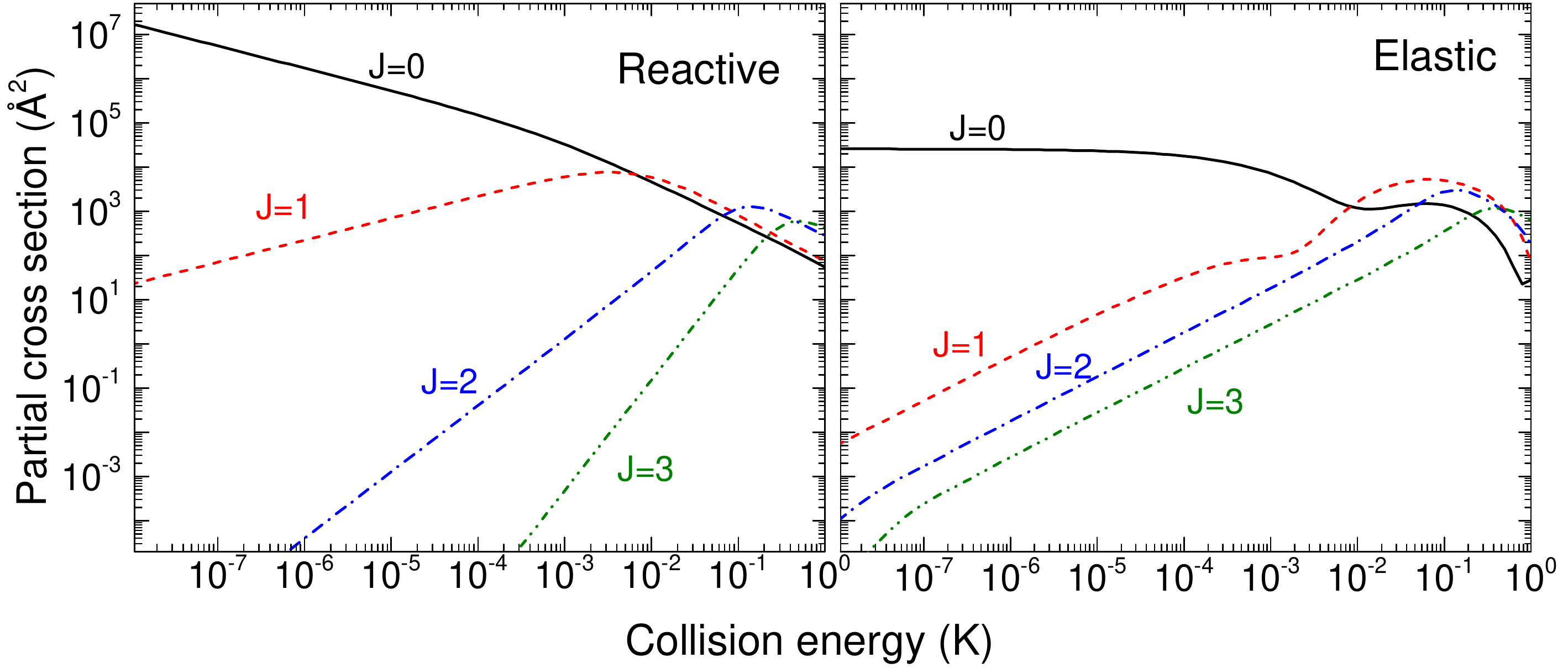}
\caption{Reaction and elastic partial cross-sections at the indicated
the $J(=l)$ values for the D$^{+}$+H$_2$ ($v$=0, $j$=0) collision
in the cold and ultracold regimes.
} \label{fig3}
\vspace{-0.5cm}
\end{center}
\end{figure}

The cross sections at much lower kinetic energies are also
shown in Fig.~\ref{fig2}. In the zero energy limit Wigner
threshold laws~\cite{Sade,Weiner} state that the elastic
and the total-loss (inelastic + reaction) cross-sections
associated to each partial wave, $l$, vary $\sim E^{2l}$
and $\sim E^{l-1/2}$, respectively. However, for a
potential with $n=4$, the threshold law for elastic
scattering becomes $\sim E$ for any
$l>0$~\cite{Sade,Weiner,Simoni}. The ultracold
cross-sections, shown in Fig.~\ref{fig3} for the four
lowest partial waves, comply with these laws (there are no
open inelastic channels). The limiting behaviors for $l=$0
are reflected in the total reaction ($\sim E^{-1/2}$) and
total elastic (constant) cross-sections in the lowest
energy region of Fig.~\ref{fig2}, where only s-wave is
open.

For $n=4$ the energy dependence of the LM coincides with
the Wigner threshold law ($\sim E^{-1/2}$). Remarkably, the
absolute values of accurate and LM cross sections in the
ultracold limit are nearly the same, $\sigma_{\rm r}
\approx 1.07 \cdot \sigma_{\rm L} (E)$, as can be seen in
Fig.~\ref{fig2}. Therefore, the reaction rate coefficient
(not shown) is practically constant in no less than eight
orders of magnitude, and small variations can be further
smoothed out with the Boltzmann averaging. In what follows,
we will try to rationalize  this classical Langevin
behavior in the ultracold regime.

Very recently, quantal versions of the LM have been
proposed~\cite{Jach:03,Gao1} under the assumption that all
the flux that reaches the SR region leads to reaction. This
situation is expected when the number of states coupled to
the incident one is so large that all the incoming flux is
irreversibly lost~\cite{Jach:03,Gao1}. In the $n=$4 case,
these {\em universal} models conclude that the zero energy
limit of $\sigma_{\rm r}$ is given by $2 \sigma_{\rm
L}(E)$, and not by $\sigma_{\rm L}(E)$ as we have
approximately obtained. Therefore our system is not
universal. The formalism in Ref.~\cite{Jach:02} is able to
deal with systems where the {\em short range reaction
probability} (SRRP), $P^{\rm re}$, is $< 1$. It provides
expressions for the complex (energy dependent) scattering
length, $\tilde{a}_{l}(k)=\alpha_{l}(k)-i \beta_{l}(k)$ in
terms of the MQDT functions (where $k$ is the relative wave
number). This allows to parametrize $\tilde{a}_{l}(k)$
using two real parameters, $y$ and $s$, together with the
mean scattering length~\cite{Griba}, $\bar{a}= (2 \mu C_4
)^{1/2}/ \hbar$ ($\approx$99.7 a$_0$ in this case).
Specifically, the dimensionless parameter $0 \le y \le 1$
characterizes the flux that is lost from the incoming
channel at SR, according to $P^{\rm re}=4y/(1+y)^2$. The
Langevin assumption or {\em universal case} corresponds to
$y=1$. The dimensionless scattering length $s=a/ \bar{a}$
is related to an entrance channel phase, where $a$ is the
s-wave scattering length corresponding to the reference
one-channel potential~\cite{Jach:02,Jach:03}. The limits $s
\rightarrow \pm \infty$ correspond to a bound state
crossing threshold.

In terms of these parameters, the small $k$ behavior of the
real and imaginary parts of the complex scattering length
for the lowest partial waves ($l$=0-3) are given
by~\footnote[5]{Equations (\ref{l0}) and (\ref{l1}) were
kindly provided by the authors of Ref.~\citenum{Jach:02};
Eq.~\eqref{l2} and Eq.~\eqref{l3} were deduced by the
authors of this work following Refs.~\citenum{Jach:02} and
~\citenum{Simoni}.}:
\begin{eqnarray}\label{l0}
\alpha_{0}(k) \to  \bar a \frac{s(1-y^2)}{1+s^2y^2} ,
\qquad \beta_{0}(k) \to  \frac{y(1+s^2)\,\bar a}{1+s^2y^2}
\end{eqnarray}
\begin{eqnarray}\label{l1}
\alpha_{1}(k)  \to  -k \bar{a}^2 \frac{ \pi}{15}, \qquad
\beta_{1}(k) \to  \frac{y(1+s^2)\, k^2 \bar{a}^3 }{9(s^2+y^2)}
\end{eqnarray}
\begin{eqnarray}\label{l2}
\alpha_{2}(k)  \to  -k \bar{a}^2 \frac{ \pi }{105} ,
\quad \beta_{2}(k)  \to
\frac{y(1+s^2)\,k^4 \bar{a}^5 }{2025(1+s^2y^2)}
\end{eqnarray}
\begin{eqnarray}\label{l3}
\alpha_{3}(k)  \to  -k \bar{a}^2 \frac{ \pi}{ 315},  \quad
\beta_{3}(k)  \to   \frac{y(1+s^2)\,k^6 \bar{a}^7}{2480625(s^2+y^2)}
\end{eqnarray}

\begin{figure}[t]
 \begin{center}
\hspace{-0cm}
\includegraphics[width=66ex]{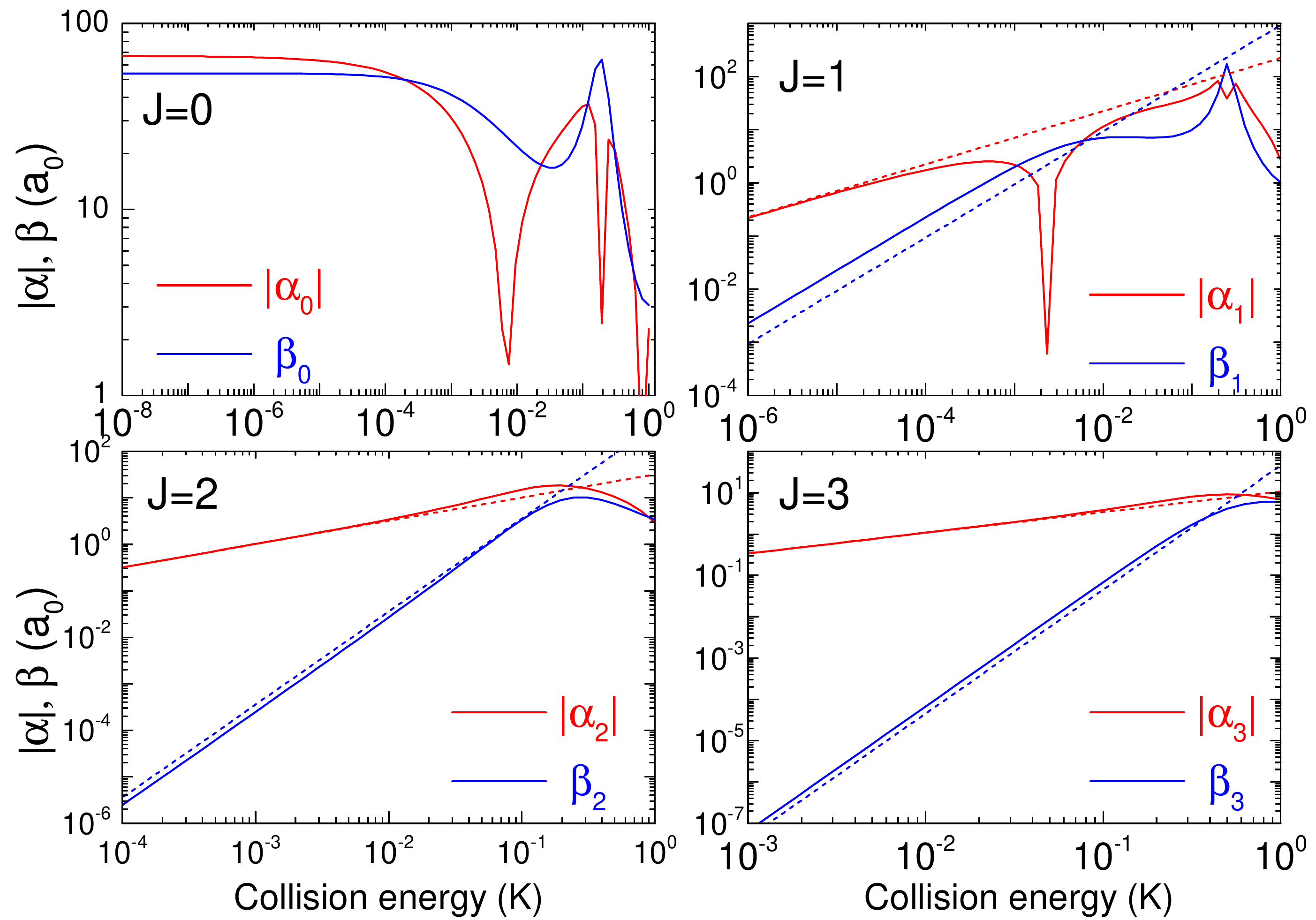}
\caption{Real and imaginary parts of the scattering length, $\tilde{a}_{l}(k)=\alpha_{l}(k)-i \beta_{l}(k)$,
obtained in the calculations for the 4 lowest partial waves. The absolute value of
$\alpha$, mostly negative, is plotted. The values for $J=$1, 2 and 3
(in continuous line) are compared with the predictions from the model in Ref.~\cite{Jach:02}
(in dashed lines), calculated using Eqs.~\eqref{l1}-\eqref{l3}, assuming for $s$ and $y$ the same values that
have been obtained for $J=0$.
} \label{fig4}
\vspace{-0.3cm}
\end{center}
\end{figure}

Our calculations yield the $S$-matrix as a function of the
energy for each total angular momentum $J$ and hence $l$
(for $j$=0), what allows to calculate directly
$\tilde{a}_{l}(k)$ using the elastic element of the
$S$-matrix~\cite{Hutson07}:
\begin{equation}
\tilde{a}_{l}(k)=\frac{1}{ik} \,\frac{1-S_{l,l}(k)}{1+S_{l,l}(k)}
\end{equation}
Fig.~\ref{fig4} depicts the energy dependence of  $\alpha$
and $\beta$ for $J=0-3$. The limiting behaviors are in
perfect agreement with the threshold laws and the power of
the dependence on $k$ in Eqns.~\eqref{l0}-\eqref{l3}.

In order to extract the model parameters $s$ and $y$ from
the calculated results, let us consider first the case
$l=0$. Using the values $\alpha_0$ and $\beta_0$ at the
lowest energy given by our calculations and solving
Eqs.~\eqref{l0} for $y$ and $s$, we obtain $y(l=0) = 0.34$
and $s(l=0) = -0.8$, which leads to a SRRP
$P^{\rm re}(l=0) = 77 \%$. The parametrization for
higher values of $l$ is not straightforward. The real part
of the scattering length, $\alpha_{l}(k)$, is independent
on $s(l)$ and $y(l)$, and with the sole expression of
$\beta_{l}(k)$ it is not possible to solve for the values
of the two parameters.

Analogously to the procedure of Ref.~\onlinecite{Jach:02},
assuming that $y$ and $s$ do not depend on $l$, we can
introduce $y(l=0)$ and $s(l=0)$ in
Eqs.~\eqref{l1}--\eqref{l3} and compare the resulting
values of $\beta_{l}$ with those obtained in the scattering
calculations. The ratios of the calculated and parametrized
values of $\beta_{l}$ are $0.4$, $1.4$, and $0.7$ for
$l=1$, $l=2$ and $l$=3, respectively. The agreement can be
considered good on average, taken into account the
oscillations of this ratio about 1. In the Ar-He($^{3}$S)
system \cite{Nar1}, where this approximation was used,
\cite{Jach:02} the reduced mass is four times bigger, what
increases the number of partial waves probably reducing the
average error. As for the the real parts of the scattering
lengths, $\alpha_l$, given by Eqs.~\eqref{l1}--\eqref{l3},
they can be directly compared with our scattering results.
The agreement (within $1\%$) is very good what can be
deemed as a test of the theory and serves to ensure the
convergence of the scattering calculations. These
expressions depend only on $\bar{a}$ (not on $s$ or $y$)
and they can be considered as really universal.

Only when SR forces act, at distances small enough as
compared to the characteristic length of the potential,
$R_4$, the dependence of the quantum defect parameter with
energy and partial wave is expected to be
negligible~\cite{Jach:02}. In the title system, $R_4$ =
$\bar{a}$ $\approx$ 100 a$_0$ is not very large, leaving
open the possibility of non constant parameters. However,
we have checked that the $\sim R^{-4}$ behavior is valid in
the region where the centrifugal barriers of the three
considered partial waves are located. Furthermore, we can
show that the dependence of $P^{\rm re}$ with $E$ and $l$
is weak. In Ref.~\cite{Jach:02}, the expression
$\sigma_{\rm r}(E) \approx P^{\rm re} \cdot \sigma_{\rm
L}(E)$ is proposed when many partial waves are open. It
connects the behavior at high energies with the value of
$P^{\rm re}$ in the ultracold regime, assuming that $y$ is
essentially constant. The fact that $\sigma_{\rm
r}(E)/\sigma_{\rm L}(E)$ has an average value of 0.78 in
the range 1-150\,K, very close to the $P^{\rm re}(l=0)$
obtained at ultracold energies, indicates a weak dependence
of $P^{\rm re}$ with $l$ and energy. An accurate
determination of $y(E,l)$ and $s(E,l)$ would require an
implementation of the MQDT fitting functions, which is out
of the scope of this work.

For $J=0$ only 3 open channels are coupled to the incident
one, what justifies a significant portion of flux returning
to the incident channel from the complex, and thus $P^{\rm
re}<1$.  Moreover, its value should reflect the statistical
balance between the reacting and returning fluxes which
results from the loss of memory at the well. We can expect
that $P^{\rm re}$, the fraction of captured flux which
reacts, is related with $P_{ \to {\rm prod}}$, the fraction
of formed complexes which decompose to give the products.
Furthermore, the expression $\sigma_{\rm r}(E) \approx
P^{\rm re} \cdot \sigma_{\rm L}(E)$, from
Ref.~\citenum{Jach:02}, has the same structure as
$\sigma_{\rm r}(E) \approx P_{ \to {\rm prod}} \cdot
\sigma_{\rm L}(E)$, obtained from the statistical
hypothesis. This establishes an {\em equivalence} between
$P^{\rm re}$ and $P_{ \to {\rm prod}}$ in the Langevin
regime. In fact, the average value of $\sigma_{\rm
r}(E)/\sigma_{\rm L}(E)$, 78\%, agrees within $10\%$ with
the statistical factor 6/7 ($\approx$86\%) at energies
$>100$\,K. This reinforces the use of statistical models
for complex--mediated systems to estimate $P^{\rm re}$, and
thus the $y$ parameter: as long as the expression
$\sigma_{\rm r}(E) \approx  P^{\rm re} \cdot \sigma_{\rm
L}(E)$ is applicable, the $P^{\rm re}$ at threshold can be
estimated by calculations of $\sigma_{\rm r}(E)/\sigma_{\rm
L}(E)$ at much higher energies (about 100\,K in our case),
where an extreme accuracy of the PES is not required and
the statistical model has proven
semiquantitative\cite{LSH-JPCA2014}. Other common dynamical
methodologies (classical and quantum) could be used to
calculate $\sigma_{\rm r}(E)/\sigma_{\rm L}(E)$ and to
extract $y$ in non-statistical systems.

In recent times, statistical methods are being revisited in
the field of cold collisions~\cite{Mayle1,Maykel}. However,
some caveats are in order if applying the statistical
ansatz in the ultracold regime. If we consider the
expression of $\beta_{0}(k)$ in Eq.~\eqref{l0}, we can
write:
\begin{equation}
P_{\rm r}^{J=0}(E)=4 k\beta_{0}(k)= 4k\bar{a} \times
\frac{y(1+s^2)}{1+s^2y^2}
\end{equation}
By comparison with the statistical factorization, $P^J_{\rm
r}(E) \approx  P^J_{\rm capt}(E) \times P_{\to {\rm
prod}}(E)$, the first factor, which is the zero-energy
limit of the $l$=0 one-channel capture
probability~\cite{Fara}, can be identified with $P^J_{\rm
capt}(E)$. Hence, the second term should be related to
$P_{\to {\rm prod}}(E)$. However, this term can be greater
than one for big values of $s$,~\cite{Jach:03} and cannot
be deemed as a probability. As $s$ is related to a phase,
and a lack of correlation between capture and decay of the
complex is implicit in the statistical model, the validity
of the ansatz is into question when approaching threshold.
Furthermore, as the energy decreases, $A(E)$, which
strictly speaking is a capture probability, will vanish
while $B(E)$ remains essentially constant. This leads to
$P_{\to {\rm prod}}(E)=1$ and thus $P_{\rm r}^{J=0}(E)=
4k\bar{a}$: {\em the zero energy limit of the statistical
ansatz is the universal case}, at least considering
one-channel captures. It is an open question how this
limitations may affect implementations of the statistical
method which consider multi-channel ({\em vs.} one-channel)
capture probabilities~\cite{Rack0}.

In summary, accurate scattering calculations have allowed
us to test the quantum theory by Jachymsky {\em et
al.}~\cite{Jach:02,Jach:03,Jach:04}. The model is able to
account for the main features of the dynamics under the
assumption of parameters independent on $l$ and $E$. On
average, the same $P^{\rm  re}$ describes the ultracold and
the Langevin regimes. We have found a link between $P^{\rm
re}$ and the statistical factor of the statistical approach
to reactions~\cite{Rack0}. Apart from physical insight,
this provides a way to calculate $y$, which characterizes
the ultracold regime, using common methodology in reaction
dynamics at much higher energies. Finally, in the light of
the model, limitations of the statistical ansatz to
describe ultracold collisions are discussed. Our work
contributes to fill the gap between the field of ultracold
atomic physics and that of quantum reaction dynamics.

The authors are greatly indebted to K. Jachymski, Andrea
Simoni and T. Gonz\'alez-Lezana for fruitful discussions
and exchange of information. The Spanish Ministries of
Science and Innovation and Economy and Competitiveness
(grants CTQ2008-02578/BQU, CSD2009-00038, and
CTQ2012-37404-C02) are gratefully acknowledged.

\end{document}